\documentclass[%
 reprint,
superscriptaddress,
 amsmath,amssymb,
 aps,
]{revtex4-2}

\usepackage{graphicx}% Include figure files
\usepackage{dcolumn}% Align table columns on decimal point
\usepackage{bm}% bold math
\usepackage{xcolor}
\usepackage{textgreek}
\begin{document}

\preprint{APS/123-QED}

\title{Terahertz spectroscopy evidence
of possible 40 K superconductivity in rhenium-doped
strontium ruthenates}
%\thanks{A footnote to the article title}%

\author{Yurii Aleshchenko}
\affiliation{V.L. Ginzburg Center for High-Temperature
Superconductivity and Quantum Materials, P.N. Lebedev Physical Institute of the Russian Academy of Sciences, 53 Leninskiy Prospekt, 119991, Moscow, Russia}

\author{Boris Gorshunov}
\affiliation{Moscow Institute of Physics and Technology (National Research University),
141700 Dolgoprudny, Moscow Region, Russia }

\author{Elena Zhukova}
\affiliation{Moscow Institute of Physics and Technology (National Research University),
141700 Dolgoprudny, Moscow Region, Russia }

\author{Andrey Muratov}
\affiliation{V.L. Ginzburg Center for High-Temperature
Superconductivity and Quantum Materials, P.N. Lebedev Physical Institute of the Russian Academy of Sciences, 53 Leninskiy Prospekt, 119991, Moscow, Russia}

\author{Alexander Dudka}
\affiliation{Shubnikov Institute of Crystallography of Federal Scientific Research
Centre \textquotedblleft Crystallography and Photonics\textquotedblright\ of
Russian Academy of Sciences, Leninskiy Prospekt 59,
119333, Moscow, Russia}

\author{Rajendra Dulal}
% \email{Second.Author@institution.edu}
\affiliation{Advanced Physics Laboratory, Institute for Quantum Studies, Chapman University, Burtonsville, MD 20866, USA}

%\collaboration{MUSO Collaboration}%\noaffiliation

\author{Serafim Teknowijoyo}
% \homepage{http://www.Second.institution.edu/~Charlie.Author}
\affiliation{Advanced Physics Laboratory, Institute for Quantum Studies, Chapman University, Burtonsville, MD 20866, USA}

\author{Sara Chahid}
\affiliation{Advanced Physics Laboratory, Institute for Quantum Studies, Chapman University, Burtonsville, MD 20866, USA}

\author{Vahan Nikoghosyan}
\affiliation{Advanced Physics Laboratory, Institute for Quantum Studies, Chapman University, Burtonsville, MD 20866, USA}

\author{Armen Gulian}
\email[Corresponding author: ]{gulian@chapman.edu}
\affiliation{Advanced Physics Laboratory, Institute for Quantum Studies, Chapman University, Burtonsville, MD 20866, USA}
% \altaffiliation[Also at ]{Physics Department, XYZ University.}%Lines break automatically or can be % forced with \\

\date{\today}% It is always \today, today,
             %  but any date may be explicitly specified

\begin{abstract}

Strontium ruthenates have many similarities with copper oxide superconductors and are of particular interest for the investigation of the mechanisms and conditions which lead to high-temperature superconductivity. We report here on multiple experimental indications of superconductivity with onset at 40~K in strontium ruthenate doped by rhenium and selenium with chlorine used as the flux. The main experimental evidence arises from terahertz spectroscopy of this material followed by AC and DC magnetization, as well as measurements of its heat capacity and magnetoresistance. Structural and morphological studies revealed the heterophase nature of this polycrystalline material as well as the changes of lattice parameters relative to the original phases. Experimental data show a higher critical temperature on the surface compared to that of the bulk of the sample. 

\end{abstract}

%\keywords{Suggested keywords}%Use showkeys class option if keyword
                              %display desired
\maketitle

%\tableofcontents

\section{Introduction}

The fascinating properties of strontium ruthenates $\mathrm{Sr}_{n+1}\mathrm{%
Ru}_{n}\mathrm{O}_{3n+1}$\textrm{\ }$(n=1,2,...,8)$ have garnered enormous
attention \cite{Armitage2019} 
since the discovery of high-temperature superconductivity in
cuprates. Superconductivity in ruthenates was found only for $n=1$ case \cite{Maeno1994}
with $T_{c}$ as high as $\mathrm{1.5K}$ \cite{Mackenzie2003}. Other representatives
of this Ruddlesden-Popper family possess peculiar magnetic properties. The
most pronounced magnetic order takes place at $n=\infty $: $\mathrm{SrRuO}%
_{3}$ is a ferromagnet with Curie temperature $\mathrm{T_{C}} \sim 165$~K; at $n=2$, $\mathrm{Sr}_{3}\mathrm{Ru}_{2}\mathrm{O}_{7}$ is an
anomalous paramagnet; at $n=3$, $\mathrm{Sr}_{4}\mathrm{Ru}_{3}\mathrm{O}%
_{10}$ is a metamagnet \cite{Mackenzie2003}. The Cooper pairing in $\mathrm{Sr}_{2}\mathrm{%
RuO}_{4}$ was well known as a textbook example of the spin-triplet state
(odd parity $\mathrm{S=1}$, see reviews \cite{Mackenzie2003,Maeno2012,Kallin2012,Liu2015} and references therein).
Recently, $\mathrm{NMR}$ spectroscopy reinvestigation \cite{Pustogow2019} has given
compelling evidence that the superconductivity in $\mathrm{Sr}_{2}\mathrm{RuO%
}_{4}$ is likely to be even parity, which unequivocally demonstrates that
research on the physical properties of ruthenates is far from being complete.

Another confirmation of this statement comes from the recent fascinating
discovery of high-temperature superconductivity in calcium ruthenate, $%
\mathrm{Ca}_{2}\mathrm{RuO}_{4}$ \cite{Nobukane2020}. The stoichiometric composition of this
material in a single-crystalline form is a Mott insulator, while single
crystals with excess oxygen are metallic above $\mathrm{160~K}$ \cite{Braden1998}. 
The excess amount of oxygen results in important crystallographic change: the lattice symmetry changes from \textit{Pbca} to $P2_1/c$ with $c-$axis extension from $11.9613$~\AA\ to $12.3719$~\AA. 
%and the crystallographic difference between them lies in the $c-$axis: 11.94~\AA ~vs. 12.35~\AA. 
Interestingly, electrically induced insulator-metal transition have been detected via
infrared nano-imaging and optical-microscopy measurements on bulk single
crystal $\mathrm{Ca}_{2}\mathrm{RuO}_{4}$ \cite{Zhang2019}. Much more drastic changes
occur when the thickness of $\mathrm{Ca}_{2}\mathrm{RuO}_{4}$ crystal is
reduced to the nanometer range: novel quantum states including
high-temperature superconductivity via resistive and magnetic measurements
at $\mathrm{64~K}$ have been observed \cite{Nobukane2020}. This remarkable finding in
micronanocrystals demonstrated how rich the superconducting phenomena in
ruthenates can be. It also invigorates the value of polycrystalline
materials (ceramics), in which the samples in \cite{Nobukane2020} were originally prepared
before subsequent sonification to obtain nanomicrocrystals.

\begin{figure}
    \centering
    \includegraphics[width=\linewidth]{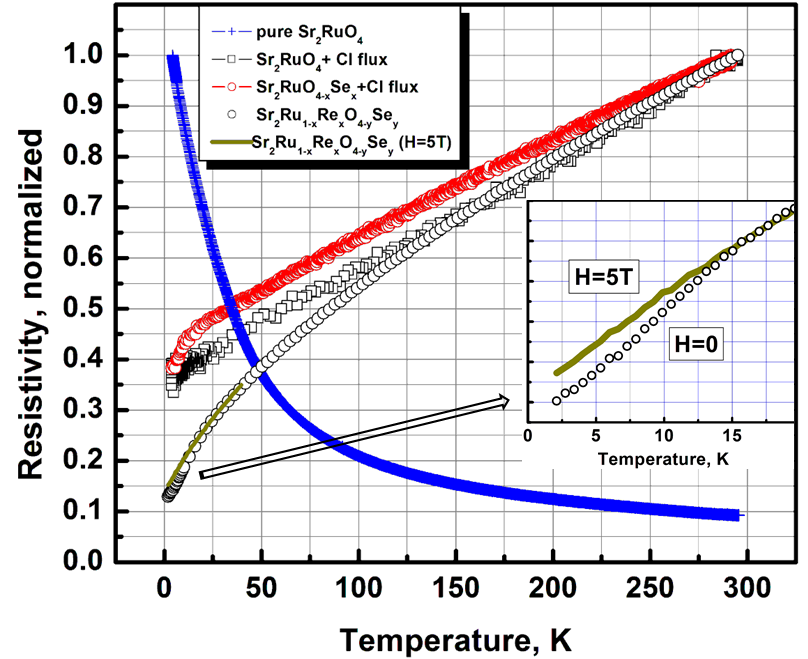}
    \caption{Drastic change in resistivity of $\mathrm{Sr}_{2}\mathrm{RuO%
}_{4}$ ceramic samples. While pure samples typically demonstrate
semiconductor-type temperature dependence, the chlorine flux and vacuum
treated samples demonstrate the so-called strange metal behavior, and chalcogen
addition introduces a downturn. Both features are intriguing (more details can
be found in \cite{Bruin2013}). The curve corresponding to Re, as well as the
modification of its resistivity in the magnetic field, are discussed in the
text.}
    \label{fig1}
\end{figure}

In this rapid communication, we present spectroscopic data in the $\mathrm{%
THz-FIR}$ range obtained for polycrystalline samples of initial
stoichiometric composition of $\mathrm{Sr}_{2}\mathrm{Ru}_{1-x}\mathrm{Re}%
_{x}\mathrm{O}_{4-y}$\textrm{Se}$_{y}$. The choice of this
composition was made in result of series of experiments in which the oxygen
was partially replaced by $\mathrm{S}$ or $\mathrm{Se}$ in presence of $%
\mathrm{Cl}$ as a flux at the synthesis of $\mathrm{Sr}_{2}\mathrm{RuO}_{4-x}%
\mathrm{S(Se)}_{x}$ and subsequent cationic substitutions for $\mathrm{Ru}$
(Fig.~\ref{fig1}).

As follows from Fig.~\ref{fig1}, typical $\mathrm{\rho }(T)$ dependence of $%
\mathrm{Sr}_{2}\mathrm{RuO}_{4}$ drastically changes into typical strange
metal behavior \cite{Bruin2013} at application of $\mathrm{Cl-}$flux and vacuum during
synthesis (details can be found in \cite{Gulian2018}). Moreover, inclination towards zero
resistivity at $T\rightarrow 0$~K appears at further substitution of
chalcogens for oxygen. Since this inclination might have been related with a
superconducting phase, we focused our efforts on deepening it using, in
particular, cationic substitutions for $\mathrm{Ru}$. The best results were
observed with $\mathrm{Re-}$ions \cite{Gulian2018}. At the time of publication \cite{Gulian2018}, the
critical temperatures in the range of $\mathrm{20-30~K}$ appeared very
unusual for ruthenates; however they were later supported by the findings of
Ref.~\cite{Nobukane2020}. Our current spectroscopic data is in support of superconducting
phase in polycrystalline $\mathrm{Sr}_{2}\mathrm{Ru}_{1-x}\mathrm{Re}_{x}%
\mathrm{O}_{4-y}$\textrm{Se}$_{y}$.

\section{Experimental details}

Details on the preparation of $\mathrm{Sr}_{2}\mathrm{Ru}_{1-x}\mathrm{Re}%
_{x}\mathrm{O}_{4-y}$\textrm{Se}$_{y}$ samples can be found in
\cite{Gulian2018}. Here, we will briefly summarize them. The precursors, $\mathrm{RuO}_{2}
$, $\mathrm{SrSe}$, $\mathrm{ReO}_{2}$, $\mathrm{SrCO}_{3}$, and $\mathrm{%
SrCl}_{2}$\textrm{\textperiodcentered }$\mathrm{6H}_{2}\mathrm{O}$ were
powdered and mixed in stoichiometric proportions. A combination of hand and
mechanical grinding and mixing was applied. The powder was calcinated at $%
\mathrm{695^{\circ}C}$ for $\mathrm{10}$ hours which incurred $\mathrm{6\%}$ of weight loss.
The calcinated powder was again powderized and heat treated in air, linearly
increasing temperature up to $\mathrm{1350
{^\circ}
C}$ and down during $\mathrm{8}$ hours with $\mathrm{25\%}$ of weight loss.
This powder was pelletized and heat treated again at $\mathrm{1350%
{{}^\circ}%
C}$\ in air for $\mathrm{5}$ hours with linear temperature increase and
decrease at a similar rate ($\mathrm{4}$ hours each, with a weight loss $%
\mathrm{\sim }\mathrm{5\%}$). For optimizing heat treatment temperatures,
powder thermogravimetry was used. Next, heat treatment of the pellet was
performed in high vacuum ($\mathrm{\sim }\mathrm{10}^{-6}~\mathrm{mbar}$%
\textrm{, }$\mathrm{650%
{{}^\circ}%
C}$\textrm{, }$\mathrm{500~\min }$). The weight of the pellet did not change
noticeably, but the resistivity became smaller. No changes in the sample's
characteristics were obtained at further vacuum heat treatments.

\begin{figure}
    \centering
    \includegraphics[width=\linewidth]{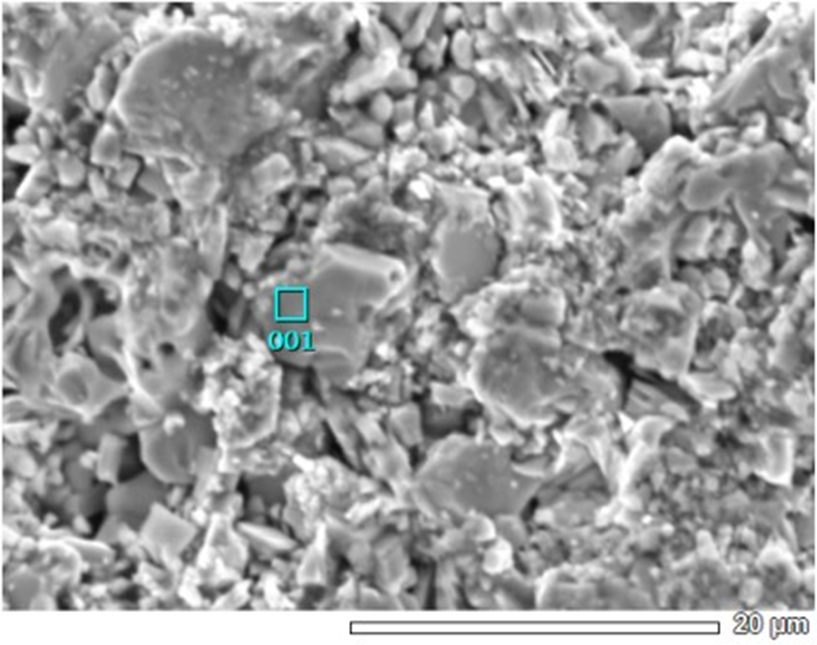}
    \caption{Polycrystalline surface morphology of the measured
sample (JEOL JCM 6000Plus SEM). The rectangle indicates one of the regions
from which the compositional data were taken.}
    \label{fig2}
\end{figure}

The crystalline structure of the sample is shown in Fig.~\ref{fig2}.
Using EDX well-matching data taken by multiple analysis of various
crystalline areas of this sample, the composition of the sample was
determined as $\mathrm{Sr}_{2.61}\mathrm{(Ru}_{1.20}\mathrm{,Re}_{0.15}%
\mathrm{)O}_{4}$. In view of $>25\%$ of weight loss at the thermal
treatment quoted above, the change of the initial stoicheometry is not
surprising (for the convenience, we will keep calling our sample $\mathrm{Sr}%
_{2}\mathrm{Ru}_{1-x}\mathrm{Re}_{x}\mathrm{O}_{4-y}$\textrm{Se}$_{y%
}$). For Se content, EDX microanalysis is not sufficiently sensitive. WDX
analysis revelas $\mathrm{Se}$ of the amount less than $0.1\%$.

For XRD structural studies (Rigaku Miniflex 600; measurement using Cu-K$%
_{\alpha} $ line in the angle interval $2\theta =3-140^\circ$ with the step $%
0.01^\circ$ and scanning rate $0.1-0.5^\circ$/s; phase content and
refinement of atomic structural models by the Panalytical
HighScore Plus 3.0e software), $\mathrm{0.5~mm}^{3}$ of sample was ground in
agath mortar. Analysis revealed the presence of three phases: $%
\mathrm{Sr}_{3}\mathrm{Ru}_{2}\mathrm{O}_{7}$ $(50-60\%)$, $\mathrm{Sr}_{2}%
\mathrm{RuO}_{4}$ $(20-30\%)$, and $\mathrm{SrRuO}_{3}$ $(10-20\%)$. The
lattice parameters for the phase $\mathrm{Sr}_{2}\mathrm{(Ru,Re)(O,Se)}_{4}$
are: $a=b=3.8745(5)$~\AA, $c=12.648(2)$~\AA%
, space group $\mathrm{I4/mmm}$. For comparison, the pure $n=1$ phase has
parameters $a=b=3.8724$~\AA, $c=12.7423$~\AA ~with the same space group \cite{Neumeier1994}. This means that the lattice
parameters of this cell are squeezed by $0.74\%$ along the $c-$axis, and
extended by $0.054\%$ within the $ab-$plane. This can be associated with an influence of the uniaxial pressure which significantly affects the $T_c$ of $\mathrm{Sr}_{2}%
\mathrm{RuO}_{4}$ \cite{Steppke2017}.
The lattice parameters for the
phase $\mathrm{Sr}_{3}\mathrm{(Ru,Re)}_{\mathrm{2}}\mathrm{(O,Se)}_{7}$ are: 
$a=b=3.8804(2)$~\AA, $c=20.664(1)$~\AA,
space group $\mathrm{I4/mmm}$. Reference data for $n=2$ \cite{Ikeda2000} are: $%
a=b=3.8872(4)$~\AA, $c=20.732(3)$~\AA. In
this case, all the lattice parameters are squeezed ($0.17\%$ for $a$, $%
0.33\%$ for $c$). 

Physical characterization of properties of this sample's magnetoresistance,
heat capacity, DC and AC magnetic susceptibility were reported in \cite{Gulian2018}
(sample \#643), and we will use them later when discussing the major topic
of this communication.

For far-infrared ($\mathrm{FIR}$) measurements, a thin (about $\mathrm{50}$~\textmu m) disk-type slice was dry cut from cylindrical sample \#643
(diameter of 4~mm) using \textrm{Princeton Scientific WS25 High Precision}
diamond-impregnated wire saw. One face of this polycrystalline slice was
carefully polished with the \textrm{Precision polishing system Allied
MultiPrep} $\mathrm{8"}$ using diamond disc with 1
~\textmu m grade to obtain a shiny, highly planar (within $\mathrm{1-2%
{{}^\circ}%
}$ accuracy) surface. The infrared reflectivity spectra were
measured near normal incidence ($\mathrm{\approx 11%
{{}^\circ}%
}$) in the spectral range of $\mathrm{40-670~cm}^{-1}$ ($\mathrm{5-83~meV}$)
at various temperatures between $\mathrm{5~K}$ and $\mathrm{300~K}$ using a
conventional Fourier transform $\mathrm{IR}$ spectrometer ($\mathrm{IFS}$ $%
\mathrm{125HR}$\textrm{,} Bruker) equipped with a liquid $\mathrm{He-}$%
cooled $\mathrm{Si-}$bolometer and a multi-layer mylar beam splitter. For $%
\mathrm{IR}$ measurements, the polished slice was mounted with the \textrm{%
STYCAST 2850ft} epoxy glue to the tip of the cone to avoid
parasitic back-reflection. A similar cone supports the gold reference mirror.
Both cones were attached to the two-position sample holder on the cold
finger of the vertical \textrm{Konti Spectro A} continuous-flow cryostat
with \textrm{TPX} windows. The design of the cryostat provides a sliding
heat exchanger with the precision positioning system controlled by stepping
motors. The possible uncertainties related to misalignments during the
taking of reference measurements, especially at low frequencies, can be
greatly reduced in relative measurements by cycling the temperature without
moving the sample \cite{Perucchi2013}. The advantage of this technique is that all
temperature-driven distortions of the optical set-up are already frozen
around\textrm{\ }$20~\mathrm{K}$, thus making it unnecessary to take a
reference measurement at every temperature in the range 5-45~K. At energies below $5%
~\mathrm{meV},$ the small size of the sample, combined with strong
oscillations that stem from standing waves between the optical elements of the
spectrometer and cryostat windows, prevent accurate measurements and set a
lower limit in our experiment.

\section{Results}

\begin{figure}
    \centering
    \includegraphics[width=\linewidth]{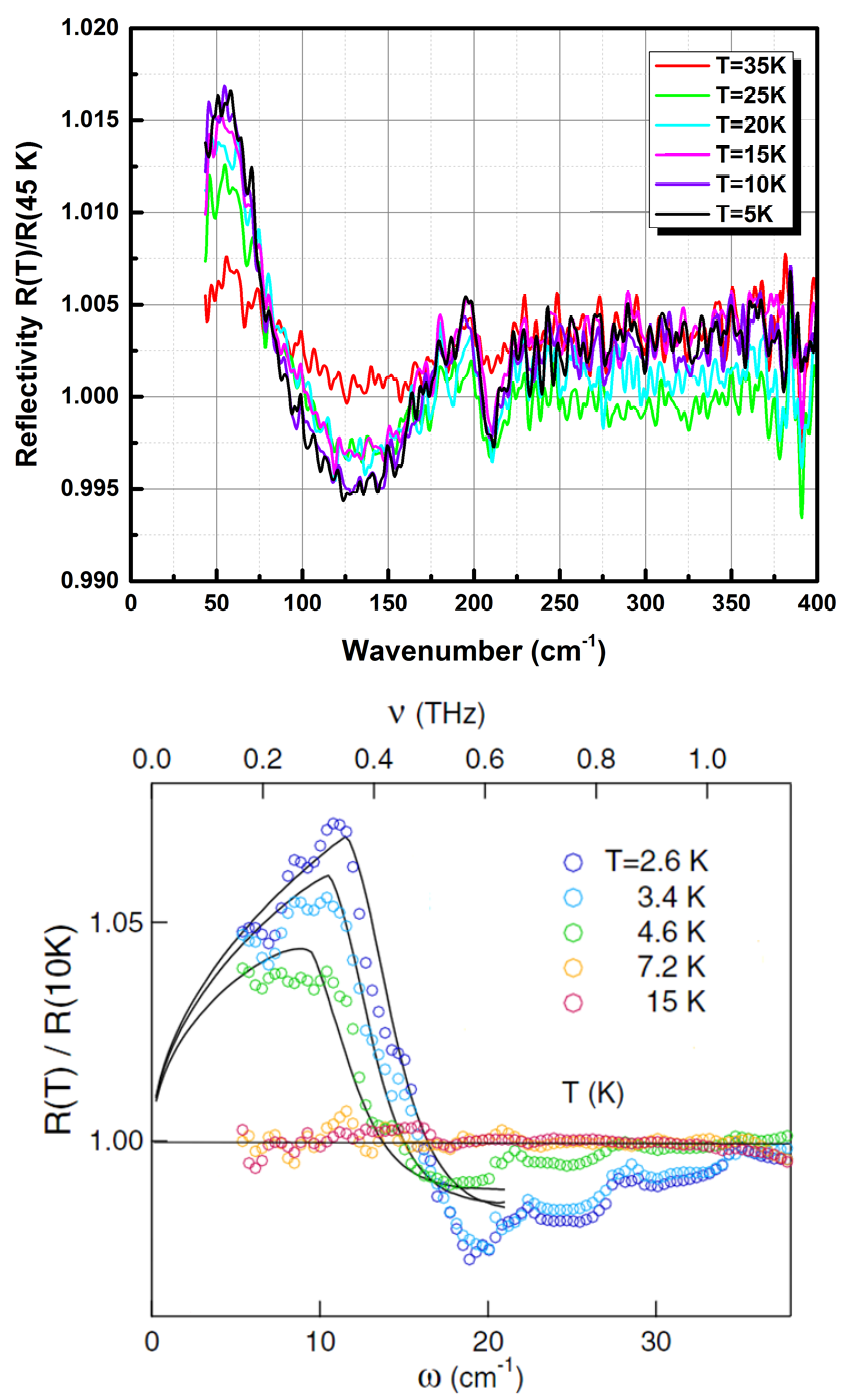}
    \caption{Reflectance of $\mathrm{Sr_{2}Ru_{1-x}Re_{x}O_{4-y}
Se_y}$ (\textit{top} panel) and of superconducting
boron-doped diamond (\textit{bottom} panel, from Ref.~\cite{Lupi2011}). In both cases, reflectance is normalized
by its normal-state value.}
    \label{fig3}
\end{figure}

Final outcome of our $\mathrm{FTIR}$ spectroscopic study is shown
in Fig.~\ref{fig3} (\textit{top} panel).
The spectral curves for $\mathrm{Sr}_{2}\mathrm{Ru}_{1-x}\mathrm{Re}_{x}%
\mathrm{O}_{4-y}$\textrm{Se}$_{y}$ were taken at $5$\textrm{, }$10$%
\textrm{, }$15$\textrm{, }$20$\textrm{, }$25$\textrm{, }$35$\textrm{\ and }$%
45~\mathrm{K}$ temperatures, then normalized by the $\mathrm{45~K}$ curve. For
wavenumbers above $220~\mathrm{cm}^{-1}$ the set of curves for different
temperatures becomes horizontal up to small vertical translations due to
noise. At the lower range of wave numbers ($40-225~\mathrm{cm}^{-1}$), one
can observe a more complex structure. The most noticeable is the first dip
which occurs within the range $75-175~\mathrm{cm}^{-1}$ and after the peak at 
$50-60~\mathrm{cm}^{-1}$. The curve corresponding to $T=35$~K has the
lowest deviation (\textit{i.e}., it does not have as deep of a dip compared
to other curves). The other curves are more packed together and reach
roughly the same peak elevation for the range $50-60~\mathrm{cm}^{-1}$.
Moreover, one can notice that the lower the temperature, the higher the
peak. This behavior closely resembles that of a typical superconductor
(\textit{e.g.} boron-doped diamond \cite{Lupi2011,Ortolani2006}) as shown in the \textit{bottom} panel of Fig~\ref{fig3}. It is important to note that for the graph in the \textit{bottom} panel
the curves correspond to temperatures both above and below the critical
temperature ($T_{c}=6$~K). In our case, $T_c$
was not known; however, one can suggest, based on the relative flatness of
the $T=35$~K curve, that $T_{c}$ should be slightly below 
\textrm{45~K}.

A second dip, though less strong in amplitude, occurs within the range $200-225%
~\mathrm{cm}^{-1}$. This dip is not accompanied by as high of a peak as the
one that was discussed above. One can theorize that this is due to another, larger gap that occurred within the range $75-175~\mathrm{cm}^{-1}$. Based on the
fact that the opening of a superconducting gap below $T_{c}$
results in the behavior shown in the \textit{bottom} panel, one can suggest
that in the \textit{top }panel at $T_{c}\sim
35$~K, two gaps of different magnitude opened up.

The dip in the case of\ the boron-doped diamond is located at $\sim 18%
~\mathrm{cm}^{-1}$. The major dip in the case of $\mathrm{Sr}_{2}\mathrm{Ru}%
_{1-x}\mathrm{Re}_{x}\mathrm{O}_{4-y}$\textrm{Se}$_{y}$ corresponds to
$\sim 125~\mathrm{cm}^{-1}$, \textit{i.e.}, the gap is by a factor of seven
larger than that of the doped diamond. This means that $T_c%
$ should be about by a factor of seven higher as well: $T_c \sim%
42$~K. Physically, the dip in reflectance corresponds to the maximum of
absorption, which takes place at the photon energy $\mathrm{\omega =2\Delta }
$ in the ``dirty" limit. Let us estimate the gap value from our data. The
wave number $\mathrm{k\sim 125~cm}^{-1}$ corresponds to photon energies $%
\mathrm{\sim 15.5~meV}$. If $T_{c}\sim 42$~K, \textit{i.e}.,
about $\mathrm{3.6~meV}$, then $2\Delta /T_{c}\sim 4.3$,
which is not far from the BCS value $3.53$; typically, higher $T_{c}
$ materials have $2\Delta /T_{c}$ ratio higher than $3.53$.

The second gap mentioned above may correspond to another phase or even have a non-superconducting
origin. Because of fluctuations, its functional form is not defined as well
as that of the larger dip. Thus, it is hard to definitively deduce the
critical temperature, as well as other relevant parameters, corresponding to
this gap. We will postpone quantitative analysis of this gap until further
exploration.

\section{Discussion}

\begin{figure}
    \centering
    \includegraphics[width=\linewidth]{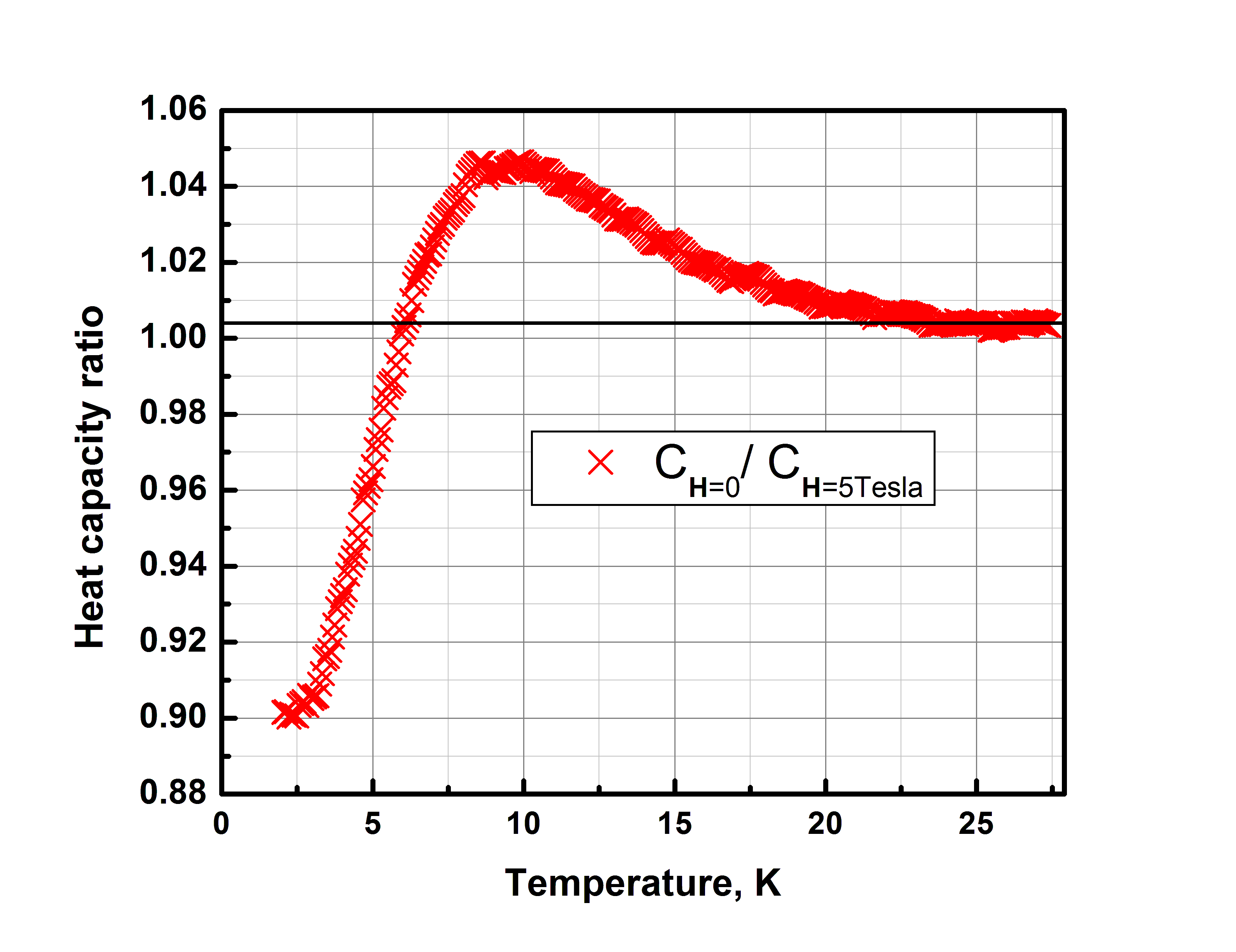}
    \caption{Broad BCS-type
singularity in heat capacity which disappears with the application of high
magnetic field.}
    \label{fig4}
\end{figure}

Let us consider how this spectroscopic result relates with other facts
reported previously on possible superconductivity in this material \cite{Gulian2018}. We
will first compare it with the heat capacity measurement which we will
replot in a more elucidating way (Fig.~\ref{fig4}).
The curve in this figure is compatible with the BCS behavior of the heat
capacity of a superconductor with a broad transition, which most
likely characterizes superconductivity in\ the heterophase $\mathrm{Sr}_{2}%
\mathrm{Ru}_{1-x}\mathrm{Re}_{x}\mathrm{O}_{4-y}$\textrm{Se}$_{y}$.
 To characterize its behavior, as shown in Fig.~\ref{fig4}, we applied a \textrm{5}~T magnetic field to the sample, which reduced the
superconducting phase volume mimicking its normal state value for the heat
capacity. One can conclude that at about $\mathrm{23~K,}$ the heat capacity
has an upturn compared to its normal value, and far below transition, it has
values lower than in its normal state (as should be expected from the
qualitative BCS pattern of superconductivity). An important question here is
why the critical temperature at this measurement is smaller by a factor of two 
than in the spectroscopic case. A possible answer can be found in the recent
results on superconductivity in calcium ruthenates \cite{Nobukane2020}. Unlike $\mathrm{60~K}$
superconductivity in $\mathrm{Ca}_{2}\mathrm{RuO}_{4}$ microcrystals,
superconductivity in bulk polycrystalline (as well as in macroscopically
large crystalline samples) is fully absent \cite{Braden1998}. If the mechanism of
superconductivity in $\mathrm{Sr}_{2}\mathrm{Ru}_{1-x}\mathrm{Re}_{x}\mathrm{%
O}_{4-y}$\textrm{Se}$_{y}$ is similar to that of $\mathrm{Ca}_{2}%
\mathrm{RuO}_{4}$ (it is hard to expect that the mechanisms are much
different!) then $\mathrm{T}_{c}$ in the bulk of $\mathrm{Sr}_{2}\mathrm{Ru}%
_{1-x}\mathrm{Re}_{x}\mathrm{O}_{4-y}$\textrm{Se}$_{y}$ pellet may
easily be lower than at the surface layer, even by a factor greater than $2$%
: the heat capacity reflects the bulk property while\ the $\mathrm{IR}$
reflectance is related with the surface layer. The micro-crystallites in the
surface layer should be relatively free from the effects of the surrounding
material.

\begin{figure}
    \centering
    \includegraphics[width=\linewidth]{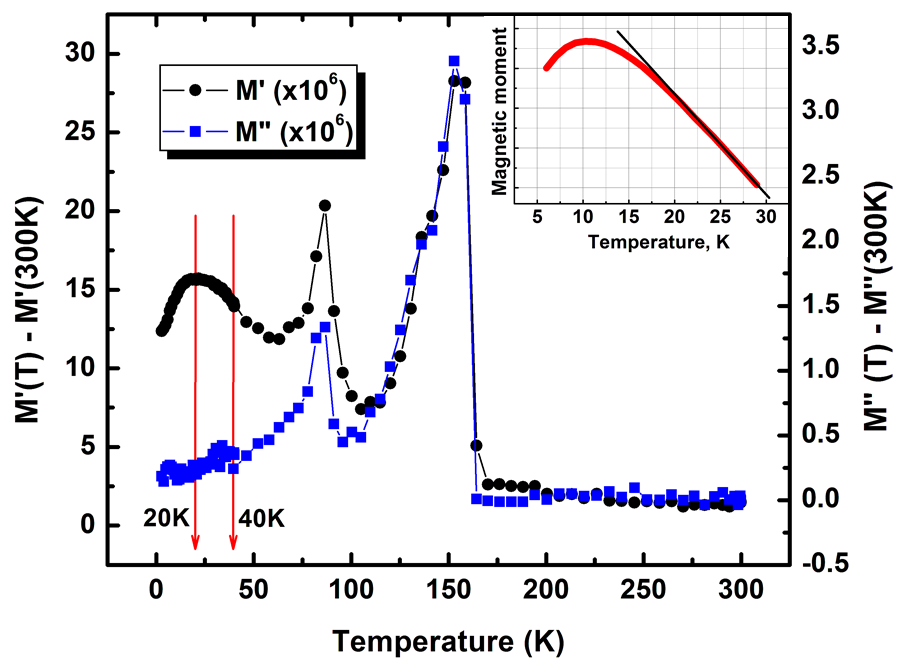}
    \caption{AC magnetic
moment $\mathrm{M=M}^{\prime }\mathrm{+iM}^{\prime \prime }$ of $%
\mathrm{Sr}_{2}\mathrm{Ru}_{1-x}\mathrm{Re}_{x}\mathrm{O}_{4-y}\mathrm{Se}_{%
y}\mathrm{.}$ Measurement with  AC field of amplitude $\mathrm{5}$ $%
\mathrm{Oe}$ and frequency $\mathrm{337~Hz}$. Inset: Magnetic moment measured
by a DC SQUID magnetometer in a 5~T field.}
    \label{fig5}
\end{figure}

The magnetization measurements support this conclusion, Fig.~\ref{fig5}.
Typically, for AC magnetic susceptibility measurements, the superconducting
transition reveals itself as a small jump at $T=T_{c}$ of the
imaginary part of the magnetic susceptibility \cite{Couach1985}. Such a jump is indeed
observable on the $\mathrm{M}''-$curve of our sample
at about $\mathrm{40~K}$, which comes close to the \textrm{FIR} data (Fig.~\ref{fig3}). After further cooling, the polycrystalline samples with intergranular
connections may have a broad hump \cite{Goldfarb1991,Civale1991,Gomory1991,Muller1991} similar to the one seen in Fig.~\ref{fig5}.
Interestingly, the major downturn of the real part of $\mathrm{M}^{\prime }$%
, as well as the downturn of the magnetic moment measured by the $\mathrm{DC}
$ magnetometer, starts at about $\mathrm{20~K}$. It appears that the relative
contribution of the surface effects is small in the case of these
quantities, similar to the heat capacity (Fig.~\ref{fig4}). These downturns may be
indicative of the Meissner effect. These curves have no hysteresis: $\mathrm{%
ZFC}$ and $\mathrm{FC}$ curves coincide, which means that the applied
magnetic field is above $H_{c1}$.

Resistive transitions in our material are incomplete, which is most likely
the result of intergranular connections in which proximitized
superconductivity is suppressed by the internal magnetic field that builds
up below $\mathrm{165~K}$ \cite{Gulian2018} due to the presence of $\mathrm{SrRuO}_{3}-$%
phase in our heterophase sample. At the same time, the resistivity downturn
becomes suppressed at the lower temperatures, as Fig.~\ref{fig1} indicates (see also
the enlarged pattern in its inset).

To complete our discussion, we should mention that features similar to the
ones which we mentioned here have been reported in the past for the
composition $\mathrm{Sr}_{3}\mathrm{Ru}_{2}\mathrm{O}_{7}$ \cite{Ikeda2000}. They were
attributed to magnetic fluctuations, similar to other cases \cite{Jarlborg1980,Mueller1970,Nakatsuji2000,Pfleiderer1997,Yoshida1998}. Leaving
aside the applicability of the magnetic fluctuations to all other facts
pointing towards superconductivity in our samples, it is hardly possible
that magnetic fluctuations would be able to quantitatively explain the
spectroscopic data presented in Section 3 (with the ratio of the gap to the
temperature of its opening being close to the BCS-value).

\section{Summary}

Terahertz spectroscopy, taken together with other observational data on $%
\mathrm{Sr}_{2}\mathrm{Ru}_{1-x}\mathrm{Re}_{x}\mathrm{O}_{4-y}\mathrm{Se}_{%
y},$ delivered indications of high temperature superconductivity
with $T_{c}^{\mathrm{onset}}\sim 40~K$ (Fig.~\ref{fig3}). This spectroscopic
value for $T_{c}$ comes close to the estimate of $T_{c}$
from the measurements of an imaginary part of the AC susceptibility (Fig.~\ref{fig5}). Both properties are likely to be determined by the surface layer of our
polycrystalline sample. Bulk characteristics, such as the heat capacity
(Fig.~\ref{fig4}) or the real part of the magnetic susceptibility (Fig.~\ref{fig5})
also point towards superconductivity and reveal themselves at lower
temperatures. That means that the bulk properties are different from the
properties of the surface layer, which sets up a bridge between our findings
and the recently discovered superconductivity at $\mathrm{60~K}$ in solitary
micronanocrystals of $\mathrm{Ca}_{2}\mathrm{RuO}_{4}$ \cite{Nobukane2020}. It is very
likely that the mechanism of superconductivity is the same in both cases.

\begin{acknowledgments}

A.D. acknowledges support by the Ministry of Science and Higher Education of
the Russian Federation (project RFMEFI62119X0035 and the State assignment of
the FSRC ``Crystallography and Photonics" RAS)
and the Shared Research Center FSRC \textquotedblleft Crystallography and
Photonics\textquotedblright\ RAS in part of X-rays diffraction study.

The work of Yu.A. and A.M. is carried out within the state assignment of the
Ministry of Science and Higher Education of the Russian Federation (theme
"Physics of high-temperature superconductors and novel quantum materials",
No. 0023-2019-0005). FIR measurements were done using research equipment of
the Shared Facilities Center at LPI.

The work of Chapman U. research team is supported by the ONR grants
N00014-16-1-2269, N00014-17-1-2972, N00014-18-1-2636 and N00014-19-1-2265.

\end{acknowledgments}

%\appendix

%\section{Appendixes}

% The \nocite command causes all entries in a bibliography to be printed out
% whether or not they are actually referenced in the text. This is appropriate
% for the sample file to show the different styles of references, but authors
% most likely will not want to use it.
%\nocite{*}

%\bibliography{references.bib}

\begin{thebibliography}{28}%
\makeatletter
\providecommand \@ifxundefined [1]{%
 \@ifx{#1\undefined}
}%
\providecommand \@ifnum [1]{%
 \ifnum #1\expandafter \@firstoftwo
 \else \expandafter \@secondoftwo
 \fi
}%
\providecommand \@ifx [1]{%
 \ifx #1\expandafter \@firstoftwo
 \else \expandafter \@secondoftwo
 \fi
}%
\providecommand \natexlab [1]{#1}%
\providecommand \enquote  [1]{``#1''}%
\providecommand \bibnamefont  [1]{#1}%
\providecommand \bibfnamefont [1]{#1}%
\providecommand \citenamefont [1]{#1}%
\providecommand \href@noop [0]{\@secondoftwo}%
\providecommand \href [0]{\begingroup \@sanitize@url \@href}%
\providecommand \@href[1]{\@@startlink{#1}\@@href}%
\providecommand \@@href[1]{\endgroup#1\@@endlink}%
\providecommand \@sanitize@url [0]{\catcode `\\12\catcode `\$12\catcode
  `\&12\catcode `\#12\catcode `\^12\catcode `\_12\catcode `\%12\relax}%
\providecommand \@@startlink[1]{}%
\providecommand \@@endlink[0]{}%
\providecommand \url  [0]{\begingroup\@sanitize@url \@url }%
\providecommand \@url [1]{\endgroup\@href {#1}{\urlprefix }}%
\providecommand \urlprefix  [0]{URL }%
\providecommand \Eprint [0]{\href }%
\providecommand \doibase [0]{https://doi.org/}%
\providecommand \selectlanguage [0]{\@gobble}%
\providecommand \bibinfo  [0]{\@secondoftwo}%
\providecommand \bibfield  [0]{\@secondoftwo}%
\providecommand \translation [1]{[#1]}%
\providecommand \BibitemOpen [0]{}%
\providecommand \bibitemStop [0]{}%
\providecommand \bibitemNoStop [0]{.\EOS\space}%
\providecommand \EOS [0]{\spacefactor3000\relax}%
\providecommand \BibitemShut  [1]{\csname bibitem#1\endcsname}%
\let\auto@bib@innerbib\@empty
%</preamble>
\bibitem [{\citenamefont {Armitage}(2019)}]{Armitage2019}%
  \BibitemOpen
  \bibfield  {author} {\bibinfo {author} {\bibfnamefont {N.~P.}\ \bibnamefont
  {Armitage}},\ }\bibfield  {title} {\bibinfo {title} {Superconductivity
  mystery turns 25},\ }\href {https://doi.org/10.1038/d41586-019-03734-7}
  {\bibfield  {journal} {\bibinfo  {journal} {Nature}\ }\textbf {\bibinfo
  {volume} {576}},\ \bibinfo {pages} {386–387} (\bibinfo {year}
  {2019})}\BibitemShut {NoStop}%
\bibitem [{\citenamefont {Maeno}\ \emph {et~al.}(1994)\citenamefont {Maeno},
  \citenamefont {Hashimoto}, \citenamefont {Yoshida}, \citenamefont
  {Nishizaki}, \citenamefont {Fujita}, \citenamefont {Bednorz},\ and\
  \citenamefont {Lichtenberg}}]{Maeno1994}%
  \BibitemOpen
  \bibfield  {author} {\bibinfo {author} {\bibfnamefont {Y.}~\bibnamefont
  {Maeno}}, \bibinfo {author} {\bibfnamefont {H.}~\bibnamefont {Hashimoto}},
  \bibinfo {author} {\bibfnamefont {K.}~\bibnamefont {Yoshida}}, \bibinfo
  {author} {\bibfnamefont {S.}~\bibnamefont {Nishizaki}}, \bibinfo {author}
  {\bibfnamefont {T.}~\bibnamefont {Fujita}}, \bibinfo {author} {\bibfnamefont
  {J.~G.}\ \bibnamefont {Bednorz}},\ and\ \bibinfo {author} {\bibfnamefont
  {F.}~\bibnamefont {Lichtenberg}},\ }\bibfield  {title} {\bibinfo {title}
  {{Superconductivity in a layered perovskite without copper}},\ }\href
  {https://doi.org/10.1038/372532a0} {\bibfield  {journal} {\bibinfo  {journal}
  {Nature}\ }\textbf {\bibinfo {volume} {372}},\ \bibinfo {pages} {532}
  (\bibinfo {year} {1994})}\BibitemShut {NoStop}%
\bibitem [{\citenamefont {Mackenzie}\ and\ \citenamefont
  {Maeno}(2003)}]{Mackenzie2003}%
  \BibitemOpen
  \bibfield  {author} {\bibinfo {author} {\bibfnamefont {A.~P.}\ \bibnamefont
  {Mackenzie}}\ and\ \bibinfo {author} {\bibfnamefont {Y.}~\bibnamefont
  {Maeno}},\ }\bibfield  {title} {\bibinfo {title} {{The superconductivity of
  ${\mathrm{Sr}}_{2}{\mathrm{RuO}}_{4}$ and the physics of spin-triplet
  pairing}},\ }\href {https://doi.org/10.1103/RevModPhys.75.657} {\bibfield
  {journal} {\bibinfo  {journal} {Rev. Mod. Phys.}\ }\textbf {\bibinfo {volume}
  {75}},\ \bibinfo {pages} {657} (\bibinfo {year} {2003})}\BibitemShut
  {NoStop}%
\bibitem [{\citenamefont {Maeno}\ \emph {et~al.}(2012)\citenamefont {Maeno},
  \citenamefont {Kittaka}, \citenamefont {Nomura}, \citenamefont {Yonezawa},\
  and\ \citenamefont {Ishida}}]{Maeno2012}%
  \BibitemOpen
  \bibfield  {author} {\bibinfo {author} {\bibfnamefont {Y.}~\bibnamefont
  {Maeno}}, \bibinfo {author} {\bibfnamefont {S.}~\bibnamefont {Kittaka}},
  \bibinfo {author} {\bibfnamefont {T.}~\bibnamefont {Nomura}}, \bibinfo
  {author} {\bibfnamefont {S.}~\bibnamefont {Yonezawa}},\ and\ \bibinfo
  {author} {\bibfnamefont {K.}~\bibnamefont {Ishida}},\ }\bibfield  {title}
  {\bibinfo {title} {{Evaluation of Spin-Triplet Superconductivity in
  Sr$_2$RuO$_4$}},\ }\href@noop {} {\bibfield  {journal} {\bibinfo  {journal}
  {Journal of the Physical Society of Japan}\ }\textbf {\bibinfo {volume}
  {81}},\ \bibinfo {pages} {011009} (\bibinfo {year} {2012})}\BibitemShut
  {NoStop}%
\bibitem [{\citenamefont {Kallin}(2012)}]{Kallin2012}%
  \BibitemOpen
  \bibfield  {author} {\bibinfo {author} {\bibfnamefont {C.}~\bibnamefont
  {Kallin}},\ }\bibfield  {title} {\bibinfo {title} {{Chiral p-wave order in
  Sr$_2$RuO$_4$}},\ }\href {https://doi.org/10.1088/0034-4885/75/4/042501}
  {\bibfield  {journal} {\bibinfo  {journal} {Reports on Progress in Physics}\
  }\textbf {\bibinfo {volume} {75}},\ \bibinfo {pages} {042501} (\bibinfo
  {year} {2012})}\BibitemShut {NoStop}%
\bibitem [{\citenamefont {Liu}\ and\ \citenamefont {Mao}(2015)}]{Liu2015}%
  \BibitemOpen
  \bibfield  {author} {\bibinfo {author} {\bibfnamefont {Y.}~\bibnamefont
  {Liu}}\ and\ \bibinfo {author} {\bibfnamefont {Z.-Q.}\ \bibnamefont {Mao}},\
  }\bibfield  {title} {\bibinfo {title} {{Unconventional superconductivity in
  Sr$_2$RuO$_4$}},\ }\href
  {https://doi.org/https://doi.org/10.1016/j.physc.2015.02.039} {\bibfield
  {journal} {\bibinfo  {journal} {Physica C: Superconductivity and its
  Applications}\ }\textbf {\bibinfo {volume} {514}},\ \bibinfo {pages} {339 }
  (\bibinfo {year} {2015})}\BibitemShut {NoStop}%
\bibitem [{\citenamefont {Pustogow}\ \emph {et~al.}(2019)\citenamefont
  {Pustogow}, \citenamefont {Luo}, \citenamefont {Chronister}, \citenamefont
  {Su}, \citenamefont {Sokolov}, \citenamefont {Jerzembeck}, \citenamefont
  {Mackenzie}, \citenamefont {Hicks}, \citenamefont {Kikugawa}, \citenamefont
  {Raghu}, \citenamefont {Bauer},\ and\ \citenamefont {Brown}}]{Pustogow2019}%
  \BibitemOpen
  \bibfield  {author} {\bibinfo {author} {\bibfnamefont {A.}~\bibnamefont
  {Pustogow}}, \bibinfo {author} {\bibfnamefont {Y.}~\bibnamefont {Luo}},
  \bibinfo {author} {\bibfnamefont {A.}~\bibnamefont {Chronister}}, \bibinfo
  {author} {\bibfnamefont {Y.-S.}\ \bibnamefont {Su}}, \bibinfo {author}
  {\bibfnamefont {D.~A.}\ \bibnamefont {Sokolov}}, \bibinfo {author}
  {\bibfnamefont {F.}~\bibnamefont {Jerzembeck}}, \bibinfo {author}
  {\bibfnamefont {A.~P.}\ \bibnamefont {Mackenzie}}, \bibinfo {author}
  {\bibfnamefont {C.~W.}\ \bibnamefont {Hicks}}, \bibinfo {author}
  {\bibfnamefont {N.}~\bibnamefont {Kikugawa}}, \bibinfo {author}
  {\bibfnamefont {S.}~\bibnamefont {Raghu}}, \bibinfo {author} {\bibfnamefont
  {E.~D.}\ \bibnamefont {Bauer}},\ and\ \bibinfo {author} {\bibfnamefont
  {S.~E.}\ \bibnamefont {Brown}},\ }\bibfield  {title} {\bibinfo {title}
  {{Constraints on the superconducting order parameter in Sr$_2$RuO$_4$ from
  oxygen-17 nuclear magnetic resonance}},\ }\href
  {https://doi.org/10.1038/s41586-019-1596-2} {\bibfield  {journal} {\bibinfo
  {journal} {Nature}\ }\textbf {\bibinfo {volume} {574}},\ \bibinfo {pages}
  {72} (\bibinfo {year} {2019})}\BibitemShut {NoStop}%
\bibitem [{\citenamefont {Nobukane}\ \emph {et~al.}(2020)\citenamefont
  {Nobukane}, \citenamefont {Yanagihara}, \citenamefont {Kunisada},
  \citenamefont {Ogasawara}, \citenamefont {Isono}, \citenamefont {Nomura},
  \citenamefont {Tanahashi}, \citenamefont {Nomura}, \citenamefont {Akiyama},\
  and\ \citenamefont {Tanda}}]{Nobukane2020}%
  \BibitemOpen
  \bibfield  {author} {\bibinfo {author} {\bibfnamefont {H.}~\bibnamefont
  {Nobukane}}, \bibinfo {author} {\bibfnamefont {K.}~\bibnamefont
  {Yanagihara}}, \bibinfo {author} {\bibfnamefont {Y.}~\bibnamefont
  {Kunisada}}, \bibinfo {author} {\bibfnamefont {Y.}~\bibnamefont {Ogasawara}},
  \bibinfo {author} {\bibfnamefont {K.}~\bibnamefont {Isono}}, \bibinfo
  {author} {\bibfnamefont {K.}~\bibnamefont {Nomura}}, \bibinfo {author}
  {\bibfnamefont {K.}~\bibnamefont {Tanahashi}}, \bibinfo {author}
  {\bibfnamefont {T.}~\bibnamefont {Nomura}}, \bibinfo {author} {\bibfnamefont
  {T.}~\bibnamefont {Akiyama}},\ and\ \bibinfo {author} {\bibfnamefont
  {S.}~\bibnamefont {Tanda}},\ }\bibfield  {title} {\bibinfo {title}
  {{Co-appearance of superconductivity and ferromagnetism in a Ca$_2$RuO$_4$
  nanofilm crystal}},\ }\href {https://doi.org/10.1038/s41598-020-60313-x}
  {\bibfield  {journal} {\bibinfo  {journal} {{Scientific Reports}}\ }\textbf
  {\bibinfo {volume} {10}},\ \bibinfo {pages} {3462} (\bibinfo {year}
  {2020})}\BibitemShut {NoStop}%
\bibitem [{\citenamefont {Braden}\ \emph {et~al.}(1998)\citenamefont {Braden},
  \citenamefont {Andr\'e}, \citenamefont {Nakatsuji},\ and\ \citenamefont
  {Maeno}}]{Braden1998}%
  \BibitemOpen
  \bibfield  {author} {\bibinfo {author} {\bibfnamefont {M.}~\bibnamefont
  {Braden}}, \bibinfo {author} {\bibfnamefont {G.}~\bibnamefont {Andr\'e}},
  \bibinfo {author} {\bibfnamefont {S.}~\bibnamefont {Nakatsuji}},\ and\
  \bibinfo {author} {\bibfnamefont {Y.}~\bibnamefont {Maeno}},\ }\bibfield
  {title} {\bibinfo {title} {{Crystal and magnetic structure of
  ${\mathrm{Ca}}_{2}{\mathrm{RuO}}_{4}:$ Magnetoelastic coupling and the
  metal-insulator transition}},\ }\href
  {https://doi.org/10.1103/PhysRevB.58.847} {\bibfield  {journal} {\bibinfo
  {journal} {Phys. Rev. B}\ }\textbf {\bibinfo {volume} {58}},\ \bibinfo
  {pages} {847} (\bibinfo {year} {1998})}\BibitemShut {NoStop}%
\bibitem [{\citenamefont {Zhang}\ \emph {et~al.}(2019)\citenamefont {Zhang},
  \citenamefont {McLeod}, \citenamefont {Han}, \citenamefont {Chen},
  \citenamefont {Bechtel}, \citenamefont {Yao}, \citenamefont {Gilbert~Corder},
  \citenamefont {Ciavatti}, \citenamefont {Tao}, \citenamefont {Aronson},
  \citenamefont {Carr}, \citenamefont {Martin}, \citenamefont {Sow},
  \citenamefont {Yonezawa}, \citenamefont {Nakamura}, \citenamefont {Terasaki},
  \citenamefont {Basov}, \citenamefont {Millis}, \citenamefont {Maeno},\ and\
  \citenamefont {Liu}}]{Zhang2019}%
  \BibitemOpen
  \bibfield  {author} {\bibinfo {author} {\bibfnamefont {J.}~\bibnamefont
  {Zhang}}, \bibinfo {author} {\bibfnamefont {A.~S.}\ \bibnamefont {McLeod}},
  \bibinfo {author} {\bibfnamefont {Q.}~\bibnamefont {Han}}, \bibinfo {author}
  {\bibfnamefont {X.}~\bibnamefont {Chen}}, \bibinfo {author} {\bibfnamefont
  {H.~A.}\ \bibnamefont {Bechtel}}, \bibinfo {author} {\bibfnamefont
  {Z.}~\bibnamefont {Yao}}, \bibinfo {author} {\bibfnamefont {S.~N.}\
  \bibnamefont {Gilbert~Corder}}, \bibinfo {author} {\bibfnamefont
  {T.}~\bibnamefont {Ciavatti}}, \bibinfo {author} {\bibfnamefont {T.~H.}\
  \bibnamefont {Tao}}, \bibinfo {author} {\bibfnamefont {M.}~\bibnamefont
  {Aronson}}, \bibinfo {author} {\bibfnamefont {G.~L.}\ \bibnamefont {Carr}},
  \bibinfo {author} {\bibfnamefont {M.~C.}\ \bibnamefont {Martin}}, \bibinfo
  {author} {\bibfnamefont {C.}~\bibnamefont {Sow}}, \bibinfo {author}
  {\bibfnamefont {S.}~\bibnamefont {Yonezawa}}, \bibinfo {author}
  {\bibfnamefont {F.}~\bibnamefont {Nakamura}}, \bibinfo {author}
  {\bibfnamefont {I.}~\bibnamefont {Terasaki}}, \bibinfo {author}
  {\bibfnamefont {D.~N.}\ \bibnamefont {Basov}}, \bibinfo {author}
  {\bibfnamefont {A.~J.}\ \bibnamefont {Millis}}, \bibinfo {author}
  {\bibfnamefont {Y.}~\bibnamefont {Maeno}},\ and\ \bibinfo {author}
  {\bibfnamefont {M.}~\bibnamefont {Liu}},\ }\bibfield  {title} {\bibinfo
  {title} {{Nano-Resolved Current-Induced Insulator-Metal Transition in the
  Mott Insulator ${\mathrm{Ca}}_{2}{\mathrm{RuO}}_{4}$}},\ }\href
  {https://doi.org/10.1103/PhysRevX.9.011032} {\bibfield  {journal} {\bibinfo
  {journal} {Phys. Rev. X}\ }\textbf {\bibinfo {volume} {9}},\ \bibinfo {pages}
  {011032} (\bibinfo {year} {2019})}\BibitemShut {NoStop}%
\bibitem [{\citenamefont {Bruin}\ \emph {et~al.}(2013)\citenamefont {Bruin},
  \citenamefont {Sakai}, \citenamefont {Perry},\ and\ \citenamefont
  {Mackenzie}}]{Bruin2013}%
  \BibitemOpen
  \bibfield  {author} {\bibinfo {author} {\bibfnamefont {J.~A.~N.}\
  \bibnamefont {Bruin}}, \bibinfo {author} {\bibfnamefont {H.}~\bibnamefont
  {Sakai}}, \bibinfo {author} {\bibfnamefont {R.~S.}\ \bibnamefont {Perry}},\
  and\ \bibinfo {author} {\bibfnamefont {A.~P.}\ \bibnamefont {Mackenzie}},\
  }\bibfield  {title} {\bibinfo {title} {{Similarity of Scattering Rates in
  Metals Showing T-Linear Resistivity}},\ }\href
  {https://doi.org/10.1126/science.1227612} {\bibfield  {journal} {\bibinfo
  {journal} {Science}\ }\textbf {\bibinfo {volume} {339}},\ \bibinfo {pages}
  {804} (\bibinfo {year} {2013})}\BibitemShut {NoStop}%
\bibitem [{\citenamefont {Gulian}\ and\ \citenamefont
  {Nikoghosyan}(2018)}]{Gulian2018}%
  \BibitemOpen
  \bibfield  {author} {\bibinfo {author} {\bibfnamefont {A.~M.}\ \bibnamefont
  {Gulian}}\ and\ \bibinfo {author} {\bibfnamefont {V.~R.}\ \bibnamefont
  {Nikoghosyan}},\ }\bibfield  {title} {\bibinfo {title} {Serendipitous vs.
  systematic search for room-temperature superconductivity},\ }\href
  {https://doi.org/10.1007/s40509-017-0143-9} {\bibfield  {journal} {\bibinfo
  {journal} {{Quantum Studies: Mathematics and Foundations}}\ }\textbf
  {\bibinfo {volume} {5}},\ \bibinfo {pages} {161} (\bibinfo {year}
  {2018})}\BibitemShut {NoStop}%
\bibitem [{\citenamefont {Neumeier}\ \emph {et~al.}(1994)\citenamefont
  {Neumeier}, \citenamefont {Hundley}, \citenamefont {Smith}, \citenamefont
  {Thompson}, \citenamefont {Allgeier}, \citenamefont {Xie}, \citenamefont
  {Yelon},\ and\ \citenamefont {Kim}}]{Neumeier1994}%
  \BibitemOpen
  \bibfield  {author} {\bibinfo {author} {\bibfnamefont {J.~J.}\ \bibnamefont
  {Neumeier}}, \bibinfo {author} {\bibfnamefont {M.~F.}\ \bibnamefont
  {Hundley}}, \bibinfo {author} {\bibfnamefont {M.~G.}\ \bibnamefont {Smith}},
  \bibinfo {author} {\bibfnamefont {J.~D.}\ \bibnamefont {Thompson}}, \bibinfo
  {author} {\bibfnamefont {C.}~\bibnamefont {Allgeier}}, \bibinfo {author}
  {\bibfnamefont {H.}~\bibnamefont {Xie}}, \bibinfo {author} {\bibfnamefont
  {W.}~\bibnamefont {Yelon}},\ and\ \bibinfo {author} {\bibfnamefont {J.~S.}\
  \bibnamefont {Kim}},\ }\bibfield  {title} {\bibinfo {title} {{Magnetic,
  thermal, transport, and structural properties of
  ${\mathrm{Sr}}_{2}$${\mathrm{RuO}}_{4+\mathrm{\ensuremath{\delta}}}$:
  Enhanced charge-carrier mass in a nearly metallic oxide}},\ }\href@noop {}
  {\bibfield  {journal} {\bibinfo  {journal} {Phys. Rev. B}\ }\textbf {\bibinfo
  {volume} {50}},\ \bibinfo {pages} {17910} (\bibinfo {year}
  {1994})}\BibitemShut {NoStop}%
\bibitem [{\citenamefont {Steppke}\ \emph {et~al.}(2017)\citenamefont
  {Steppke}, \citenamefont {Zhao}, \citenamefont {Barber}, \citenamefont
  {Scaffidi}, \citenamefont {Jerzembeck}, \citenamefont {Rosner}, \citenamefont
  {Gibbs}, \citenamefont {Maeno}, \citenamefont {Simon}, \citenamefont
  {Mackenzie},\ and\ \citenamefont {Hicks}}]{Steppke2017}%
  \BibitemOpen
  \bibfield  {author} {\bibinfo {author} {\bibfnamefont {A.}~\bibnamefont
  {Steppke}}, \bibinfo {author} {\bibfnamefont {L.}~\bibnamefont {Zhao}},
  \bibinfo {author} {\bibfnamefont {M.~E.}\ \bibnamefont {Barber}}, \bibinfo
  {author} {\bibfnamefont {T.}~\bibnamefont {Scaffidi}}, \bibinfo {author}
  {\bibfnamefont {F.}~\bibnamefont {Jerzembeck}}, \bibinfo {author}
  {\bibfnamefont {H.}~\bibnamefont {Rosner}}, \bibinfo {author} {\bibfnamefont
  {A.~S.}\ \bibnamefont {Gibbs}}, \bibinfo {author} {\bibfnamefont
  {Y.}~\bibnamefont {Maeno}}, \bibinfo {author} {\bibfnamefont {S.~H.}\
  \bibnamefont {Simon}}, \bibinfo {author} {\bibfnamefont {A.~P.}\ \bibnamefont
  {Mackenzie}},\ and\ \bibinfo {author} {\bibfnamefont {C.~W.}\ \bibnamefont
  {Hicks}},\ }\bibfield  {title} {\bibinfo {title} {{Strong peak in T$_c$ of
  Sr$_2$RuO$_4$ under uniaxial pressure}},\ }\href@noop {} {\bibfield
  {journal} {\bibinfo  {journal} {Science}\ }\textbf {\bibinfo {volume} {355}}
  (\bibinfo {year} {2017})}\BibitemShut {NoStop}%
\bibitem [{\citenamefont {Ikeda}\ \emph {et~al.}(2000)\citenamefont {Ikeda},
  \citenamefont {Maeno}, \citenamefont {Nakatsuji}, \citenamefont {Kosaka},\
  and\ \citenamefont {Uwatoko}}]{Ikeda2000}%
  \BibitemOpen
  \bibfield  {author} {\bibinfo {author} {\bibfnamefont {S.-I.}\ \bibnamefont
  {Ikeda}}, \bibinfo {author} {\bibfnamefont {Y.}~\bibnamefont {Maeno}},
  \bibinfo {author} {\bibfnamefont {S.}~\bibnamefont {Nakatsuji}}, \bibinfo
  {author} {\bibfnamefont {M.}~\bibnamefont {Kosaka}},\ and\ \bibinfo {author}
  {\bibfnamefont {Y.}~\bibnamefont {Uwatoko}},\ }\bibfield  {title} {\bibinfo
  {title} {{Ground state in
  ${\mathrm{Sr}}_{3}{\mathrm{Ru}}_{2}{\mathrm{O}}_{7}:$ Fermi liquid close to a
  ferromagnetic instability}},\ }\href
  {https://doi.org/10.1103/PhysRevB.62.R6089} {\bibfield  {journal} {\bibinfo
  {journal} {Phys. Rev. B}\ }\textbf {\bibinfo {volume} {62}},\ \bibinfo
  {pages} {R6089} (\bibinfo {year} {2000})}\BibitemShut {NoStop}%
\bibitem [{\citenamefont {Perucchi}\ \emph {et~al.}(2013)\citenamefont
  {Perucchi}, \citenamefont {Baldassarre}, \citenamefont {Joseph},
  \citenamefont {Lupi}, \citenamefont {Lee}, \citenamefont {Eom}, \citenamefont
  {Jiang}, \citenamefont {Weiss}, \citenamefont {Hellstrom},\ and\
  \citenamefont {Dore}}]{Perucchi2013}%
  \BibitemOpen
  \bibfield  {author} {\bibinfo {author} {\bibfnamefont {A.}~\bibnamefont
  {Perucchi}}, \bibinfo {author} {\bibfnamefont {L.}~\bibnamefont
  {Baldassarre}}, \bibinfo {author} {\bibfnamefont {B.}~\bibnamefont {Joseph}},
  \bibinfo {author} {\bibfnamefont {S.}~\bibnamefont {Lupi}}, \bibinfo {author}
  {\bibfnamefont {S.}~\bibnamefont {Lee}}, \bibinfo {author} {\bibfnamefont
  {C.~B.}\ \bibnamefont {Eom}}, \bibinfo {author} {\bibfnamefont
  {J.}~\bibnamefont {Jiang}}, \bibinfo {author} {\bibfnamefont {J.~D.}\
  \bibnamefont {Weiss}}, \bibinfo {author} {\bibfnamefont {E.~E.}\ \bibnamefont
  {Hellstrom}},\ and\ \bibinfo {author} {\bibfnamefont {P.}~\bibnamefont
  {Dore}},\ }\bibfield  {title} {\bibinfo {title} {{Transmittance and
  reflectance measurements at terahertz frequencies on a superconducting
  BaFe$_{1.84}$Co$_{0.16}$As$_2$ ultrathin film: an analysis of the optical
  gaps in the Co-doped BaFe$_2$As$_2$ pnictide}},\ }\href
  {https://doi.org/10.1140/epjb/e2013-30964-y} {\bibfield  {journal} {\bibinfo
  {journal} {The European Physical Journal B}\ }\textbf {\bibinfo {volume}
  {86}},\ \bibinfo {pages} {274} (\bibinfo {year} {2013})}\BibitemShut
  {NoStop}%
\bibitem [{\citenamefont {Lupi}(2011)}]{Lupi2011}%
  \BibitemOpen
  \bibfield  {author} {\bibinfo {author} {\bibfnamefont {S.}~\bibnamefont
  {Lupi}},\ }\bibfield  {title} {\bibinfo {title} {{Terahertz Spectroscopy of
  Novel Superconductors}},\ }\href {https://doi.org/10.1155/2011/816906}
  {\bibfield  {journal} {\bibinfo  {journal} {Advances in Condensed Matter
  Physics}\ }\textbf {\bibinfo {volume} {2011}},\ \bibinfo {pages} {816906}
  (\bibinfo {year} {2011})}\BibitemShut {NoStop}%
\bibitem [{\citenamefont {Ortolani}\ \emph {et~al.}(2006)\citenamefont
  {Ortolani}, \citenamefont {Lupi}, \citenamefont {Baldassarre}, \citenamefont
  {Schade}, \citenamefont {Calvani}, \citenamefont {Takano}, \citenamefont
  {Nagao}, \citenamefont {Takenouchi},\ and\ \citenamefont
  {Kawarada}}]{Ortolani2006}%
  \BibitemOpen
  \bibfield  {author} {\bibinfo {author} {\bibfnamefont {M.}~\bibnamefont
  {Ortolani}}, \bibinfo {author} {\bibfnamefont {S.}~\bibnamefont {Lupi}},
  \bibinfo {author} {\bibfnamefont {L.}~\bibnamefont {Baldassarre}}, \bibinfo
  {author} {\bibfnamefont {U.}~\bibnamefont {Schade}}, \bibinfo {author}
  {\bibfnamefont {P.}~\bibnamefont {Calvani}}, \bibinfo {author} {\bibfnamefont
  {Y.}~\bibnamefont {Takano}}, \bibinfo {author} {\bibfnamefont
  {M.}~\bibnamefont {Nagao}}, \bibinfo {author} {\bibfnamefont
  {T.}~\bibnamefont {Takenouchi}},\ and\ \bibinfo {author} {\bibfnamefont
  {H.}~\bibnamefont {Kawarada}},\ }\bibfield  {title} {\bibinfo {title}
  {{Low-Energy Electrodynamics of Superconducting Diamond}},\ }\href
  {https://doi.org/10.1103/PhysRevLett.97.097002} {\bibfield  {journal}
  {\bibinfo  {journal} {Phys. Rev. Lett.}\ }\textbf {\bibinfo {volume} {97}},\
  \bibinfo {pages} {097002} (\bibinfo {year} {2006})}\BibitemShut {NoStop}%
\bibitem [{\citenamefont {Couach}\ \emph {et~al.}(1985)\citenamefont {Couach},
  \citenamefont {Khoder},\ and\ \citenamefont {Monnier}}]{Couach1985}%
  \BibitemOpen
  \bibfield  {author} {\bibinfo {author} {\bibfnamefont {M.}~\bibnamefont
  {Couach}}, \bibinfo {author} {\bibfnamefont {A.}~\bibnamefont {Khoder}},\
  and\ \bibinfo {author} {\bibfnamefont {F.}~\bibnamefont {Monnier}},\
  }\bibfield  {title} {\bibinfo {title} {Study of superconductors by a.c.
  susceptibility},\ }\href
  {https://doi.org/https://doi.org/10.1016/0011-2275(85)90190-0} {\bibfield
  {journal} {\bibinfo  {journal} {Cryogenics}\ }\textbf {\bibinfo {volume}
  {25}},\ \bibinfo {pages} {695 } (\bibinfo {year} {1985})}\BibitemShut
  {NoStop}%
\bibitem [{\citenamefont {Goldfarb}\ \emph {et~al.}(1991)\citenamefont
  {Goldfarb}, \citenamefont {Lelental},\ and\ \citenamefont
  {Thompson}}]{Goldfarb1991}%
  \BibitemOpen
  \bibfield  {author} {\bibinfo {author} {\bibfnamefont {R.~B.}\ \bibnamefont
  {Goldfarb}}, \bibinfo {author} {\bibfnamefont {M.}~\bibnamefont {Lelental}},\
  and\ \bibinfo {author} {\bibfnamefont {C.~A.}\ \bibnamefont {Thompson}},\
  }\bibinfo {title} {{Alternating-Field Susceptometry and Magnetic
  Susceptibility of Superconductors}},\ in\ \href
  {https://doi.org/10.1007/978-1-4899-2379-0_3} {\emph {\bibinfo {booktitle}
  {Magnetic Susceptibility of Superconductors and Other Spin Systems}}},\
  \bibinfo {editor} {edited by\ \bibinfo {editor} {\bibfnamefont {R.~A.}\
  \bibnamefont {Hein}}, \bibinfo {editor} {\bibfnamefont {T.~L.}\ \bibnamefont
  {Francavilla}},\ and\ \bibinfo {editor} {\bibfnamefont {D.~H.}\ \bibnamefont
  {Liebenberg}}}\ (\bibinfo  {publisher} {Springer US},\ \bibinfo {address}
  {Boston, MA},\ \bibinfo {year} {1991})\ pp.\ \bibinfo {pages}
  {49--80}\BibitemShut {NoStop}%
\bibitem [{\citenamefont {Civale}\ \emph {et~al.}(1991)\citenamefont {Civale},
  \citenamefont {Worthington}, \citenamefont {Krusin-Elbaum},\ and\
  \citenamefont {Holtzberg}}]{Civale1991}%
  \BibitemOpen
  \bibfield  {author} {\bibinfo {author} {\bibfnamefont {L.}~\bibnamefont
  {Civale}}, \bibinfo {author} {\bibfnamefont {T.~K.}\ \bibnamefont
  {Worthington}}, \bibinfo {author} {\bibfnamefont {L.}~\bibnamefont
  {Krusin-Elbaum}},\ and\ \bibinfo {author} {\bibfnamefont {F.}~\bibnamefont
  {Holtzberg}},\ }\bibinfo {title} {{Nonlinear A.C. Susceptibility Response
  Near the Irreversibility Line}},\ in\ \href
  {https://doi.org/10.1007/978-1-4899-2379-0_15} {\emph {\bibinfo {booktitle}
  {Magnetic Susceptibility of Superconductors and Other Spin Systems}}},\
  \bibinfo {editor} {edited by\ \bibinfo {editor} {\bibfnamefont {R.~A.}\
  \bibnamefont {Hein}}, \bibinfo {editor} {\bibfnamefont {T.~L.}\ \bibnamefont
  {Francavilla}},\ and\ \bibinfo {editor} {\bibfnamefont {D.~H.}\ \bibnamefont
  {Liebenberg}}}\ (\bibinfo  {publisher} {Springer US},\ \bibinfo {address}
  {Boston, MA},\ \bibinfo {year} {1991})\ pp.\ \bibinfo {pages}
  {313--332}\BibitemShut {NoStop}%
\bibitem [{\citenamefont {G{\"o}m{\"o}ry}(1991)}]{Gomory1991}%
  \BibitemOpen
  \bibfield  {author} {\bibinfo {author} {\bibfnamefont {F.}~\bibnamefont
  {G{\"o}m{\"o}ry}},\ }\bibinfo {title} {{Responses of High T$_c$
  Superconductors to Various Combinations of AC and DC Magnetic Fields}},\ in\
  \href {https://doi.org/10.1007/978-1-4899-2379-0_14} {\emph {\bibinfo
  {booktitle} {Magnetic Susceptibility of Superconductors and Other Spin
  Systems}}},\ \bibinfo {editor} {edited by\ \bibinfo {editor} {\bibfnamefont
  {R.~A.}\ \bibnamefont {Hein}}, \bibinfo {editor} {\bibfnamefont {T.~L.}\
  \bibnamefont {Francavilla}},\ and\ \bibinfo {editor} {\bibfnamefont {D.~H.}\
  \bibnamefont {Liebenberg}}}\ (\bibinfo  {publisher} {Springer US},\ \bibinfo
  {address} {Boston, MA},\ \bibinfo {year} {1991})\ pp.\ \bibinfo {pages}
  {289--311}\BibitemShut {NoStop}%
\bibitem [{\citenamefont {M{\"u}ller}(1991)}]{Muller1991}%
  \BibitemOpen
  \bibfield  {author} {\bibinfo {author} {\bibfnamefont {K.-H.}\ \bibnamefont
  {M{\"u}ller}},\ }\bibinfo {title} {{Detailed Theory of the Magnetic Response
  of High-Temperature Ceramic Superconductors}},\ in\ \href
  {https://doi.org/10.1007/978-1-4899-2379-0_10} {\emph {\bibinfo {booktitle}
  {Magnetic Susceptibility of Superconductors and Other Spin Systems}}},\
  \bibinfo {editor} {edited by\ \bibinfo {editor} {\bibfnamefont {R.~A.}\
  \bibnamefont {Hein}}, \bibinfo {editor} {\bibfnamefont {T.~L.}\ \bibnamefont
  {Francavilla}},\ and\ \bibinfo {editor} {\bibfnamefont {D.~H.}\ \bibnamefont
  {Liebenberg}}}\ (\bibinfo  {publisher} {Springer US},\ \bibinfo {address}
  {Boston, MA},\ \bibinfo {year} {1991})\ pp.\ \bibinfo {pages}
  {229--250}\BibitemShut {NoStop}%
\bibitem [{\citenamefont {Jarlborg}\ and\ \citenamefont
  {Freeman}(1980)}]{Jarlborg1980}%
  \BibitemOpen
  \bibfield  {author} {\bibinfo {author} {\bibfnamefont {T.}~\bibnamefont
  {Jarlborg}}\ and\ \bibinfo {author} {\bibfnamefont {A.~J.}\ \bibnamefont
  {Freeman}},\ }\bibfield  {title} {\bibinfo {title} {{Magnetism and
  superconductivity in $C15$ compounds from self-consistent band
  calculations}},\ }\href {https://doi.org/10.1103/PhysRevB.22.2332} {\bibfield
   {journal} {\bibinfo  {journal} {Phys. Rev. B}\ }\textbf {\bibinfo {volume}
  {22}},\ \bibinfo {pages} {2332} (\bibinfo {year} {1980})}\BibitemShut
  {NoStop}%
\bibitem [{\citenamefont {Mueller}\ \emph {et~al.}(1970)\citenamefont
  {Mueller}, \citenamefont {Freeman}, \citenamefont {Dimmock},\ and\
  \citenamefont {Furdyna}}]{Mueller1970}%
  \BibitemOpen
  \bibfield  {author} {\bibinfo {author} {\bibfnamefont {F.~M.}\ \bibnamefont
  {Mueller}}, \bibinfo {author} {\bibfnamefont {A.~J.}\ \bibnamefont
  {Freeman}}, \bibinfo {author} {\bibfnamefont {J.~O.}\ \bibnamefont
  {Dimmock}},\ and\ \bibinfo {author} {\bibfnamefont {A.~M.}\ \bibnamefont
  {Furdyna}},\ }\bibfield  {title} {\bibinfo {title} {{Electronic Structure of
  Palladium}},\ }\href {https://doi.org/10.1103/PhysRevB.1.4617} {\bibfield
  {journal} {\bibinfo  {journal} {Phys. Rev. B}\ }\textbf {\bibinfo {volume}
  {1}},\ \bibinfo {pages} {4617} (\bibinfo {year} {1970})}\BibitemShut
  {NoStop}%
\bibitem [{\citenamefont {Nakatsuji}\ and\ \citenamefont
  {Maeno}(2000)}]{Nakatsuji2000}%
  \BibitemOpen
  \bibfield  {author} {\bibinfo {author} {\bibfnamefont {S.}~\bibnamefont
  {Nakatsuji}}\ and\ \bibinfo {author} {\bibfnamefont {Y.}~\bibnamefont
  {Maeno}},\ }\bibfield  {title} {\bibinfo {title} {{Quasi-Two-Dimensional Mott
  Transition System
  ${\mathrm{Ca}}_{2\ensuremath{-}\mathit{x}}{\mathrm{Sr}}_{\mathit{x}}{\mathrm{RuO}}_{4}$}},\
  }\href {https://doi.org/10.1103/PhysRevLett.84.2666} {\bibfield  {journal}
  {\bibinfo  {journal} {Phys. Rev. Lett.}\ }\textbf {\bibinfo {volume} {84}},\
  \bibinfo {pages} {2666} (\bibinfo {year} {2000})}\BibitemShut {NoStop}%
\bibitem [{\citenamefont {Pfleiderer}\ \emph {et~al.}(1997)\citenamefont
  {Pfleiderer}, \citenamefont {McMullan}, \citenamefont {Julian},\ and\
  \citenamefont {Lonzarich}}]{Pfleiderer1997}%
  \BibitemOpen
  \bibfield  {author} {\bibinfo {author} {\bibfnamefont {C.}~\bibnamefont
  {Pfleiderer}}, \bibinfo {author} {\bibfnamefont {G.~J.}\ \bibnamefont
  {McMullan}}, \bibinfo {author} {\bibfnamefont {S.~R.}\ \bibnamefont
  {Julian}},\ and\ \bibinfo {author} {\bibfnamefont {G.~G.}\ \bibnamefont
  {Lonzarich}},\ }\bibfield  {title} {\bibinfo {title} {{Magnetic quantum phase
  transition in MnSi under hydrostatic pressure}},\ }\href
  {https://doi.org/10.1103/PhysRevB.55.8330} {\bibfield  {journal} {\bibinfo
  {journal} {Phys. Rev. B}\ }\textbf {\bibinfo {volume} {55}},\ \bibinfo
  {pages} {8330} (\bibinfo {year} {1997})}\BibitemShut {NoStop}%
\bibitem [{\citenamefont {Yoshida}\ \emph {et~al.}(1998)\citenamefont
  {Yoshida}, \citenamefont {Nakamura}, \citenamefont {Goko}, \citenamefont
  {Fujita}, \citenamefont {Maeno}, \citenamefont {Mori},\ and\ \citenamefont
  {NishiZaki}}]{Yoshida1998}%
  \BibitemOpen
  \bibfield  {author} {\bibinfo {author} {\bibfnamefont {K.}~\bibnamefont
  {Yoshida}}, \bibinfo {author} {\bibfnamefont {F.}~\bibnamefont {Nakamura}},
  \bibinfo {author} {\bibfnamefont {T.}~\bibnamefont {Goko}}, \bibinfo {author}
  {\bibfnamefont {T.}~\bibnamefont {Fujita}}, \bibinfo {author} {\bibfnamefont
  {Y.}~\bibnamefont {Maeno}}, \bibinfo {author} {\bibfnamefont
  {Y.}~\bibnamefont {Mori}},\ and\ \bibinfo {author} {\bibfnamefont
  {S.}~\bibnamefont {NishiZaki}},\ }\bibfield  {title} {\bibinfo {title}
  {{Electronic crossover in the highly anisotropic normal state of
  ${\mathrm{Sr}}_{2}{\mathrm{RuO}}_{4}$ from pressure effects on electrical
  resistivity}},\ }\href {https://doi.org/10.1103/PhysRevB.58.15062} {\bibfield
   {journal} {\bibinfo  {journal} {Phys. Rev. B}\ }\textbf {\bibinfo {volume}
  {58}},\ \bibinfo {pages} {15062} (\bibinfo {year} {1998})}\BibitemShut
  {NoStop}%
\end{thebibliography}

%apsrev4-2.bst 2019-01-14 (MD) hand-edited version of apsrev4-1.bst
%Control: key (0)
%Control: author (8) initials jnrlst
%Control: editor formatted (1) identically to author
%Control: production of article title (0) allowed
%Control: page (0) single
%Control: year (1) truncated
%Control: production of eprint (0) enabled
\providecommand{\noopsort}[1]{}\providecommand{\singleletter}[1]{#1}%

\end{document}